\documentclass[a4paper]{jpconf}
\usepackage{graphicx}
\usepackage{footmisc}

\begin{document}
\title{Unitary neutron matter in the on-shell
  limit~\footnote{Talk given by V.S.T. at XXXVII RTFNB, 37th
    Brazilian Workshop on Nuclear Physics, 8-12 September 2014,
    Maresias, SP, Brazil}}

\author{E. Ruiz Arriola$^1$, S. Szpigel$^2$ and V. S. Tim\'oteo$^3$}

\address{
               $^1$Departamento de Fisica At\'omica, Molecular y Nuclear, Universidad de Granada\\
                       E-18071, Granada, Andalucia, Espa\~na\\
               $^2$Centro de R\'adio-Astronomia, Universidade Presbiteriana Mackenzie\\
                       01302-907, São Paulo, SP, Brasil\\
               $^3$Grupo de \'Optica e Modelagem Num\'erica - GOMNI, Faculdade de Tecnologia - FT\\
               Universidade Estadual de Campinas - UNICAMP, 
                       13484-332, Limeira, SP , Brasil\\
         }

\ead{earriola@ugr.es (ERA)} 
\ead{szpigel@mackenzie.br (SS)}
\ead{varese@ft.unicamp.br (VST)}

\begin{abstract}
We compute the Bertsch parameter for neutron matter by using
nucleon-nucleon interactions that are fully diagonal in momentum
space.  We analyze the on-shell limit with the similarity
renormalization group and compare the results for a simple separable toy model 
to realistic calculations with high precision $NN$ potentials.
\end{abstract}

\section{Introduction}

About fifteen years ago George Bertsch proposed the following problem
\cite{pb1999}: what would be the ground state properties of a
many-body system composed of spin-$\frac 1 2$ fermions interacting via
a short-range contact interaction with an infinitely large scattering
length?

The two-body scattering amplitude can be written as
\begin{equation}
{\cal T}_2 (k,k) \propto  \frac{1}{[k\cot\delta -ik]} \; ,
\end{equation}
and at low energies, $k\cot\delta$ can be described by the effective range expansion
\begin{equation}
k\cot\delta = -\frac 1 \alpha_0 + \frac 1 2 r_0 k^2 + \cdots \; ,
\end{equation}
where $\alpha_0$ is the scattering length and $r_0$ is the effective
range. The unitary limit corresponds to $\alpha_0\to\infty$ and
$r_0\to0$. In this limit, $k\cot\delta\to0$ and the scattering
amplitude is then reduced to
\begin{equation}
{\cal T}_2 (k,k) \propto \frac i k  \; .
\end{equation}
and thus the cross section in the $S$-wave saturates the unitarity
bound $\sigma = 4\pi / k^2$. Therefore the two-body scattering
amplitude becomes completely scale invariant and as a consequence the
energy per particle for the ground state of such a system would be
given by
\begin{equation}
\varepsilon = \xi \times \varepsilon_{FG} \; ,
\end{equation}
where 
\begin{equation}
\varepsilon_{FG} =  \frac 3 5 \frac{\hbar^2 k^2_F}{2m}
\end{equation}
is the energy per particle of the free (non-interacting) Fermi Gas and
$\xi$ has been called Bertsch parameter. In fact, Bertsch wanted to
know the sign of $\xi$.  If $\xi>0$, the system is a gas with positive
pressure, and if $\xi<0$ the system collapses since the pressure is
negative.

\begin{figure}[h]
\begin{center}
\includegraphics[width=30pc]{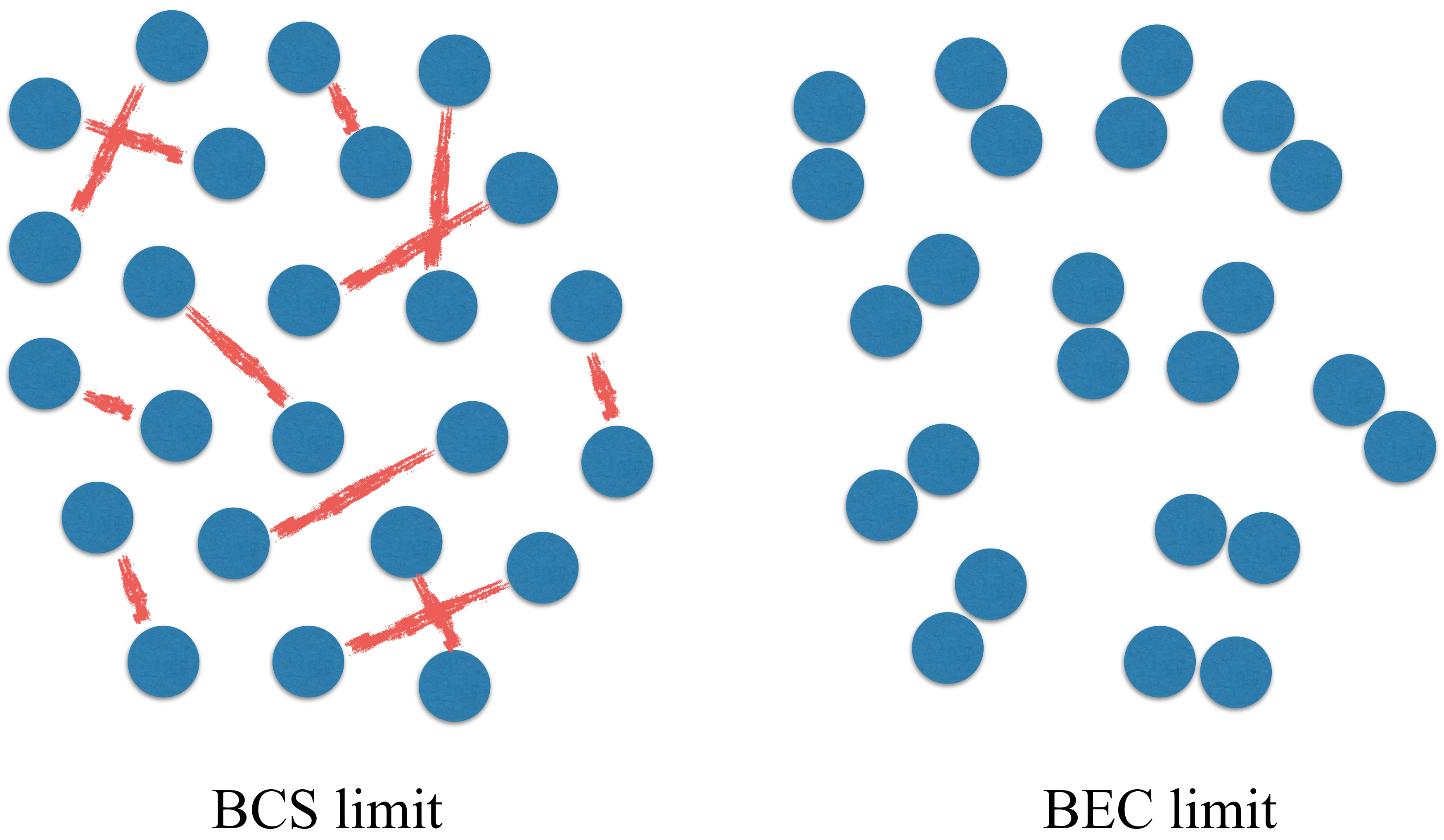}
\end{center}
\caption{Pictorical representation of the BCS and BEC limits of a
  many-fermion system.  The blurred red lines indicate the weak
  pairing. Adapted from Living Rev. Relativity {\bf 11}, 10 (2008).}
\label{fig1}
\end{figure}

Interesting many-body physics emerges when the interaction is close to
the unitarity limit. In particular, many-body fermionic systems behave
almost like a perfect fluid and may exhibit both BCS crossover and
Bose-Einstein condensation which are, in fact, distinct limits of a
common phenomenon. If the scattering length is small and negative, the
interaction is weakly attractive and the system is in the BCS limit
and there are overlapping loosely bound pairs. If the scattering
length is large and negative, the interaction is strongly attractive
and the system is in the BEC limit where the fermions form deeply
bound pairs \cite{ch2008} (see Fig. \ref{fig1}).

Both neutron matter at low densities and ultra-cold atoms close to a
Feshbach resonance are strongly interacting fermionic systems and 
present large pairing gaps when measured in units of the Fermi energy
\cite{cgg2012}. Monte Carlo simulations of a superfluid Fermi gas with
an attractive short-range two-body interaction in the unitary limit
(infinite scattering length and zero effective range) estimate the
energy per particle of neutron matter in 44\% of the energy of the
free Fermi gas and the pairing gap to be about twice the energy per
particle \cite{ccps2003}.

The $S$-wave interaction between two neutrons is very attractive
(almost enough to produce a $nn$ bound state) and has a significantly
large (and negative) scattering length, $- \alpha_0^{nn}= 18.5 {\rm
  fm} \gg r_0^{nn} = 2.7 {\rm fm}$. Hence, neutron matter at low
densities is a system with features close to the unitary
limit. Another reason to study neutron matter at low densities is that
both superconductivity and superfluidity in fermonic systems are
manifestations of quantum coherence at a macroscopic level. An {\it ab
  initio} calculation of a Fermi gas in the unitary limit shows that,
at $T = 0.2~E_F$, the viscosity is close to the lower limit for a
perfect fluid \cite{wmd2012}.

\section{On-shell interactions from the SRG and the Bertsch parameter}

In this work we want to study neutron matter at the unitary limit with
on-shell interactions obtained by evolving the nuclear force with the
similarity renormalization group (SRG) towards the infrared region of the
similarity cutoff $\lambda$. The SRG has been widely applied to
nuclear structure calculations and the nuclear many-body problem
\cite{bfs2010,fh2013}. The technique is based on a flow equation which
for the nucleon-nucleon interaction and the Wilson generator reads
\begin{eqnarray}
\frac{dV_s(p,p')}{ds} = -(p^2-p'^2)^2~V_s(p,p') + \int_0^\infty dq~q^2(p^2+p'^2-2q^2) V_s(p,q)~V_s(q,p') \; ,
\end{eqnarray}
where the flow parameter $s$ is usually written in terms of the
so-called similarity cutoff $\lambda$ (which has dimension of
momentum) as $s = \lambda^{-4}$. The potential before evolution
(initial) corresponds to $s=0$ or $\lambda=\infty$ and the matrix
elements of the evolved potential are denoted as $V_\lambda(p,p')$.
The unitarity of the transformation ensures that all observables
computed with $V_\lambda(p,p')$ are exactly the same as the
observables computed with the non-evolved initial potential
$V_\infty(p,p')$. This applies in particular to phase-shifts, which do
not depend on $\lambda$.

Recently, we have developed techniques in nuclear physics in order to
study the infrared fixed-point of the SRG by pushing the evolution
towards the on-shell limit $\lambda \to
0$~\cite{ast2014plb,ast2014fbs,rst2014aop} and have found an elegant
an simple way to determine phase shifts from fully diagonal
interactions in momentum space complying with isospectrality and
Levinson's theorem \cite{rst2014plb}.

A simple $S$-wave  gaussian separable potential toy model
allows to carry studies with a moderate numerical effort 
\begin{equation}
V(p,p') = C~\exp \left[ - \frac{1}{L^2} \left( p^2 + p'^2 \right)
  \right] \; ,
\end{equation}
where the parameters $C$ and $L$ are obtained by fitting the
scattering length and the effective range.  The $nn$ interaction
cannot be measured directly, but since the $nn$ and $np$ interaction
in the $^1S_0$ channel have similar (and large) scattering lengths,
$\alpha_0^{nn} = -18.5~{\rm fm}$ and $\alpha_0^{np} = -23.7~{\rm fm}$,
we use the $np$ phase-shifts to access how well the toy model
describes the nuclear force in the $S$-waves. This gives 
$C= -1.916~{\rm fm} $ and $L=1.2~{\rm fm}^{-1}$.

At similarity cutoffs close to $\lambda \sim 1~{\rm fm}^{-1}$ the flow
equation becomes extremely stiff so that it is nearly impossible to
study the infrared limit of the similarity cutoff, $\lambda\to0$, if
the potential has a long tail in momentum space, which is the case for
high precision nucleon-nucleon potentials. This is the reason why we
have constructed the toy model since it gives good qualitative results
for the nucleon-nucleon $S$-waves but has a short tail in momentum
space, allowing the SRG evolution towards the infrared region of the
similarity cutoff with a moderate numerical effort. The fully diagonal
on-shell interaction at $\lambda=0$ is obtained by using the energy
shift prescription of Ref. \cite{rst2014plb}.

\section{Bertsch parameter}

Here we apply the toy model to compute the Bertsch parameter in the
infrared region of the similarity cutoff with different grid sizes.
We also compute the Bertsch parameter with high precision
nucleon-nucleon potentials to compare them to the results from the toy
potential.

The Bertsch parameter is the ratio between the total energy of a
system of interacting fermions in the unitary limit and the energy of
a free Fermi gas:
\begin{eqnarray}
\xi_\lambda (k_F) = \frac{T(k_F)+V_\lambda (k_F)}{T(k_F)} = 1 + \frac{V_\lambda (k_F)}{T(k_F)} \; .
\end{eqnarray}
The kinetic energy in neutron matter is given by
\begin{eqnarray}
T(k_F) = \frac{3 k_F^2}{10 m_n} \; ,
\end{eqnarray}
where $m_n$ is the neutron mass and $k_F$ is the Fermi momentum. The
potential energy can be obtained in the Hartree-Fock approximation and
is due to the toy interaction in the $^1S_0$ channel:
\begin{eqnarray}
V_\lambda (k_F)= \frac{4}{m_n} \frac{2}{\pi} \int_0^{k_F} dk~k^2  
 \left( 1 - \frac{3k}{2k_F} + \frac{k^3}{2k_F^3} \right) ~ V^{^1S_0}_\lambda (k,k) \; .
\end{eqnarray}
For realistic high precision interactions, there are also
contributions from higher partial waves and the energy per particle
can be written as
\begin{eqnarray}
\varepsilon (k_F) = \frac{3k^2_F}{10m_n} &+& \frac{4}{m_n} \frac{2}{\pi} \int_0^{k_F} dk~k^2  
 \left( 1 - \frac{3k}{2k_F} + \frac{k^3}{2k_F^3} \right) \nonumber \\ 
 &\times& \left[  V^{^1S_0}_\lambda (k,k) + 9V^{^3P_c}_\lambda (k,k) + 5V^{^1D_2}_\lambda (k,k) 
 + 21V^{^3F_c}_\lambda (k,k) + 9V^{^1G_4}_\lambda (k,k) \right] \; , \nonumber \\ 
\end{eqnarray}
where $^3P_c$ and $^3F_c$ are linear combinations of the $P$ and $F$ waves, which are given
explicitly in Ref. \cite{ast2012}.

\section{Numerical results}

\begin{figure}[h]
\begin{center}
\includegraphics[width=17pc]{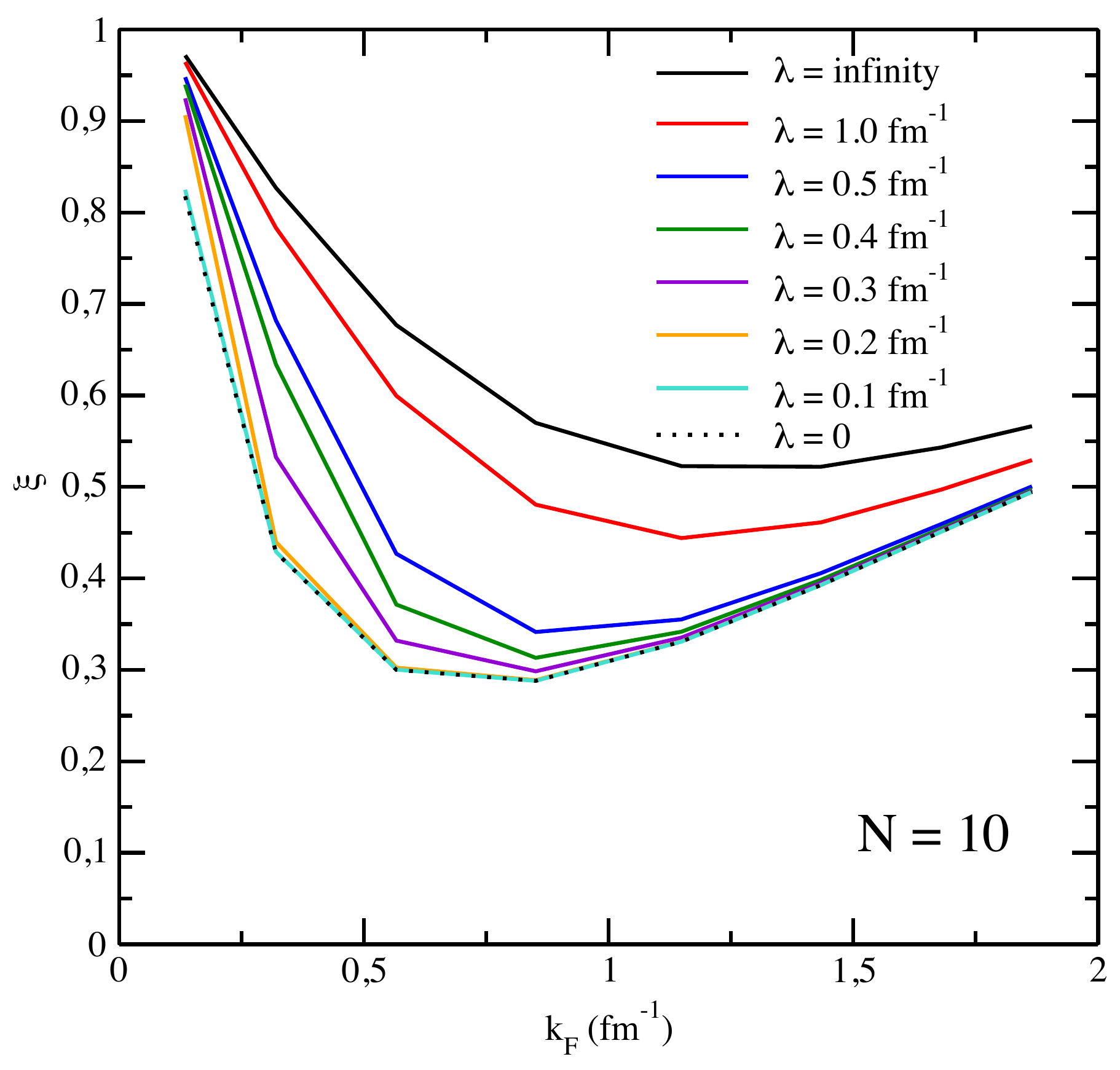} \hspace{0.5cm}
\includegraphics[width=17pc]{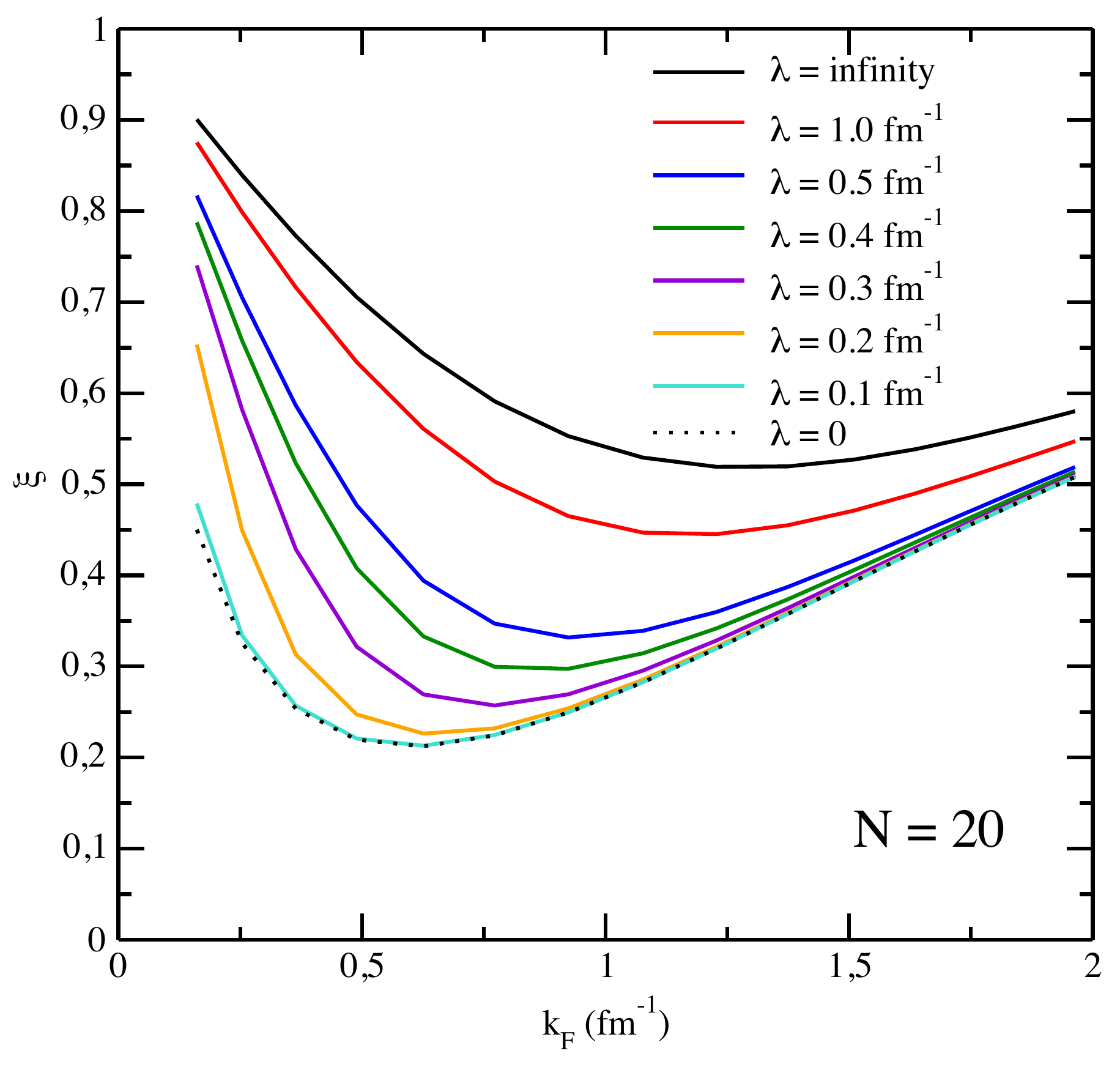} \\
\includegraphics[width=17pc]{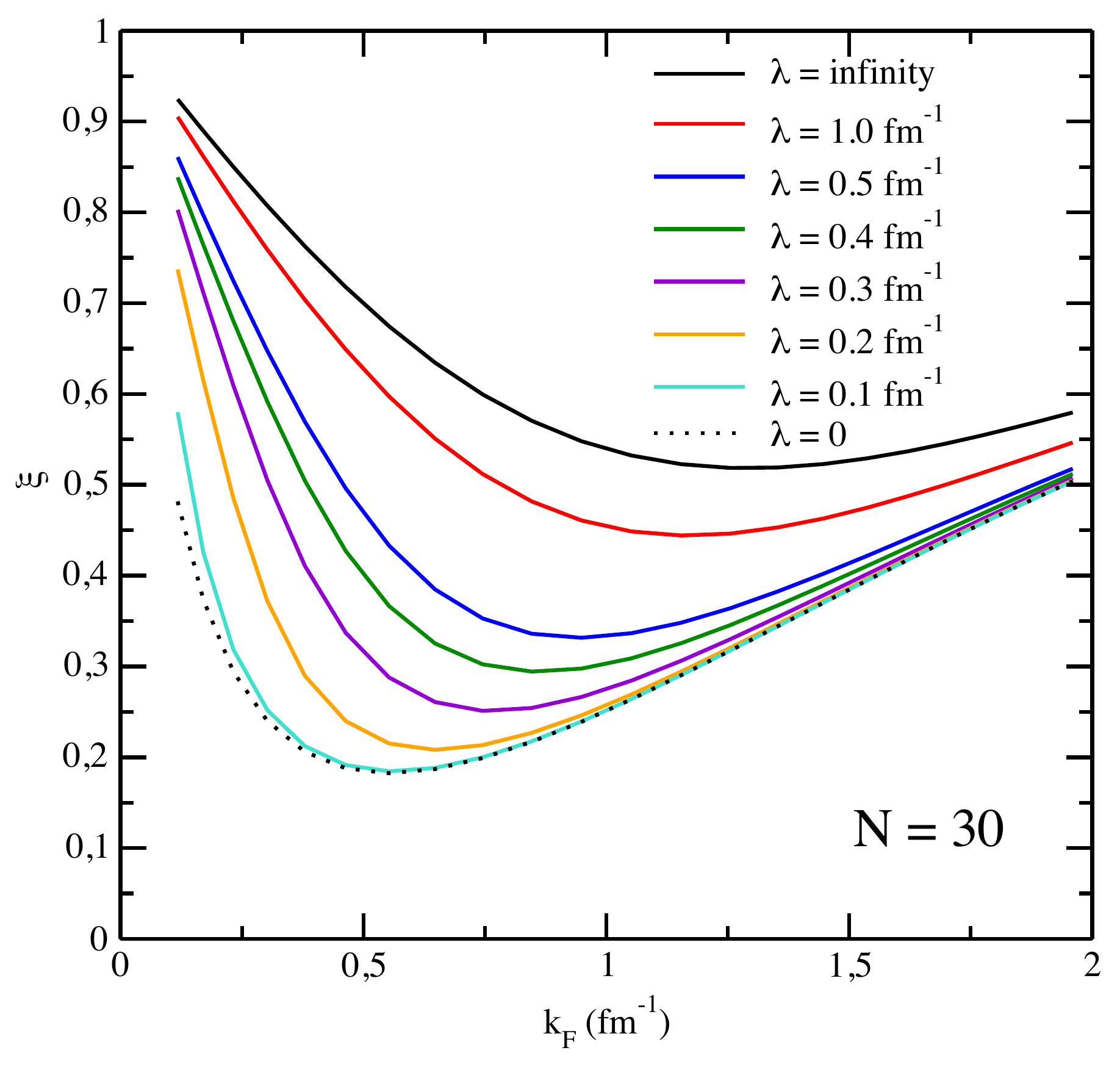} \hspace{0.5cm}
\includegraphics[width=17pc]{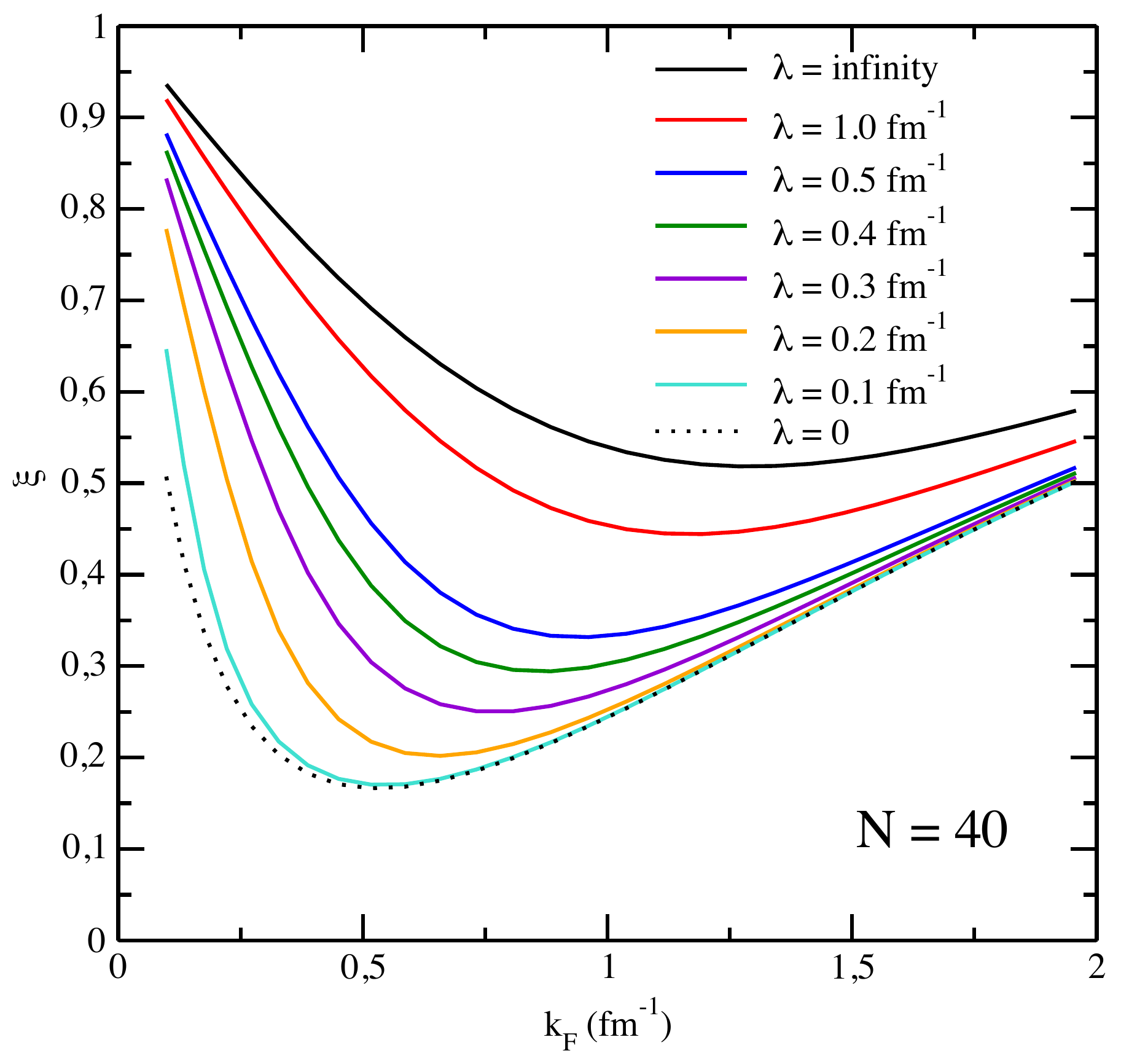} \\
\includegraphics[width=17pc]{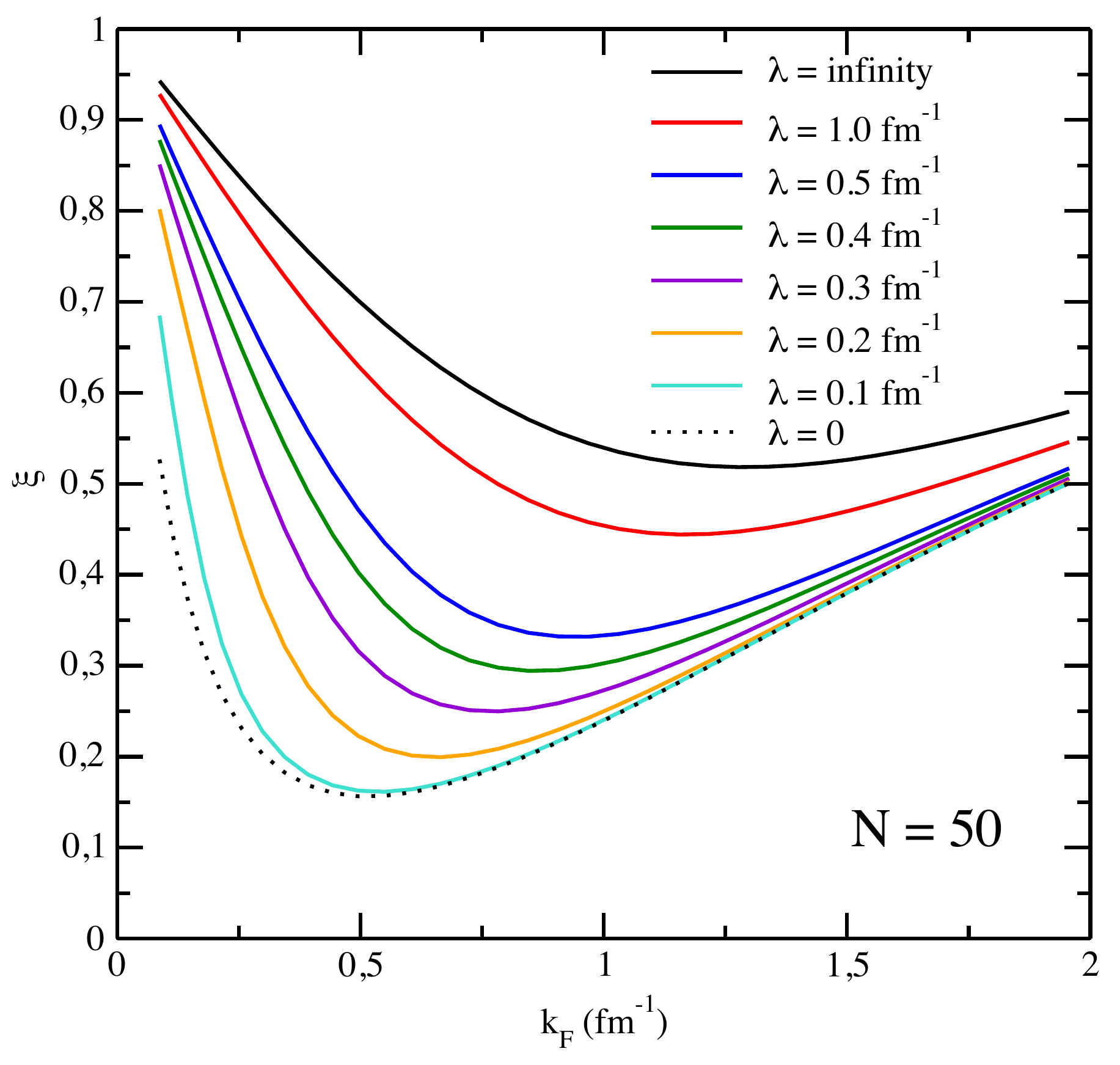} \hspace{0.5cm}
\includegraphics[width=17pc]{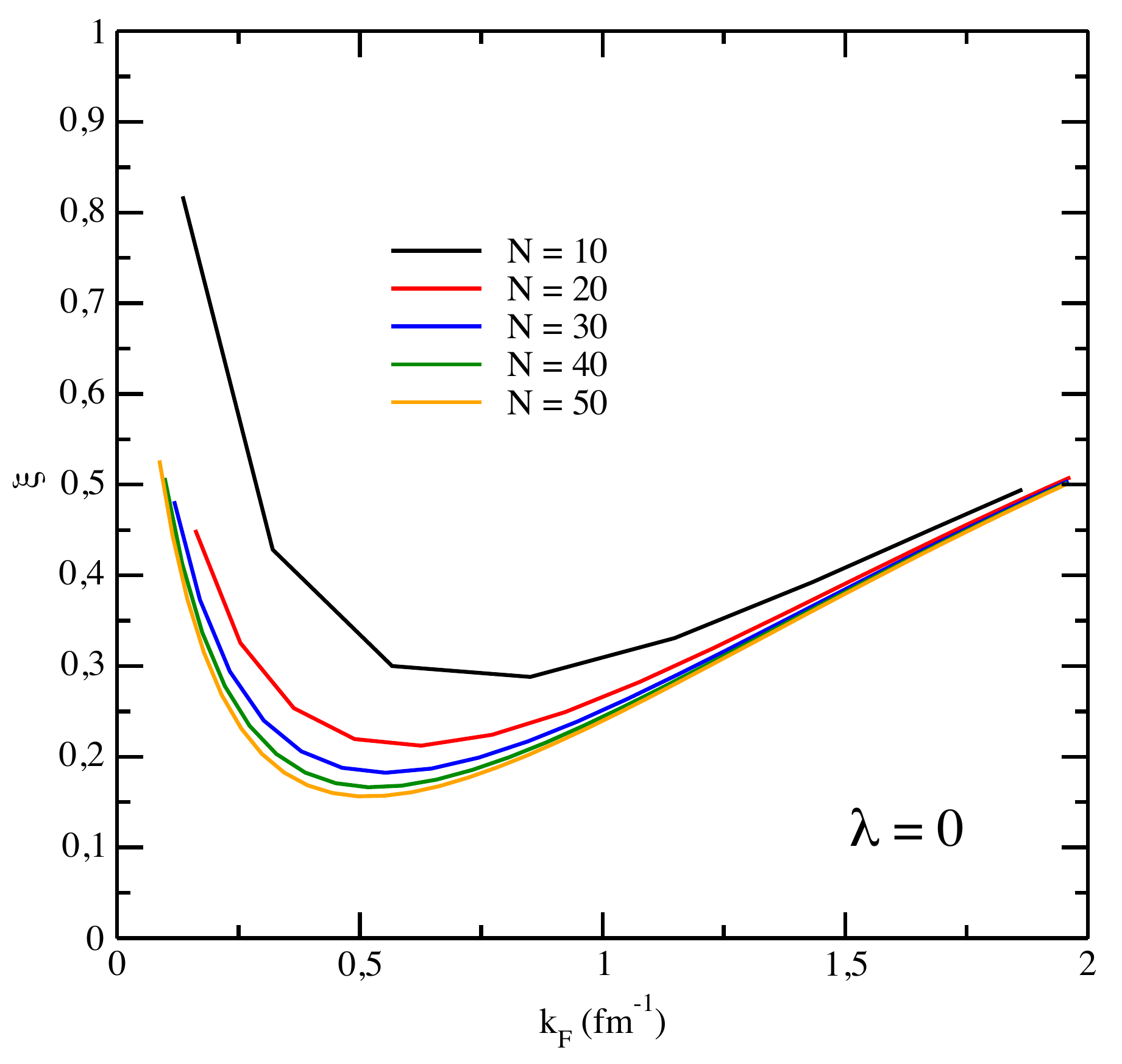} \\
\end{center}
\caption{Bertsch parameter from the $S$-wave toy model for several values of the similarity cutoff $\lambda$ (Wilson generator) 
and different number of grid points $N$. The right panel in the last row shows the grid dependence in the limit $\lambda=0$. 
The on-shell interaction $V_{\lambda=0}(k)$ was obtained by applying the eigenvalue method of Ref. \cite{rst2014plb}.}
\label{fig2}
\end{figure}

In Fig. \ref{fig2} we show the Bertsch parameter as a function of the
Fermi momentum computed with the toy model for several values of the
similarity cutoff, mostly in the infrared region, for different number
of grid points. For similarity cutoffs from infinity down to $0.5~{\rm
  fm}^{-1}$ the number of grid points almost do not affect the
results. However, for smaller values of $\lambda$ the number of grid
points start to change the results and in the limit $\lambda=0$ the
difference is huge as can be seen in the last panel of
Fig. \ref{fig2}.

\begin{figure}[t]
\begin{center}
\includegraphics[width=12pc]{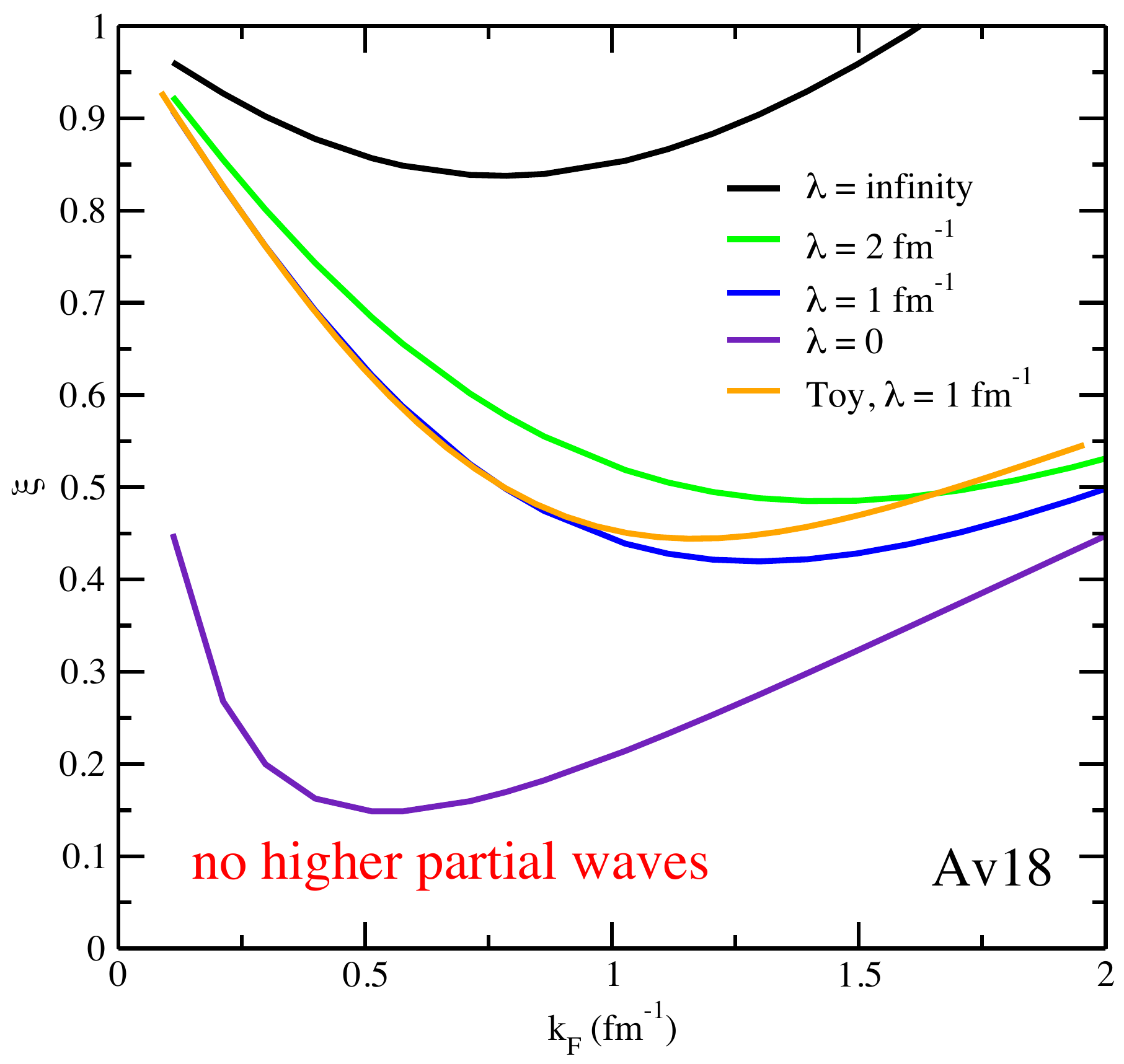}\hspace{1cm}
\includegraphics[width=12pc]{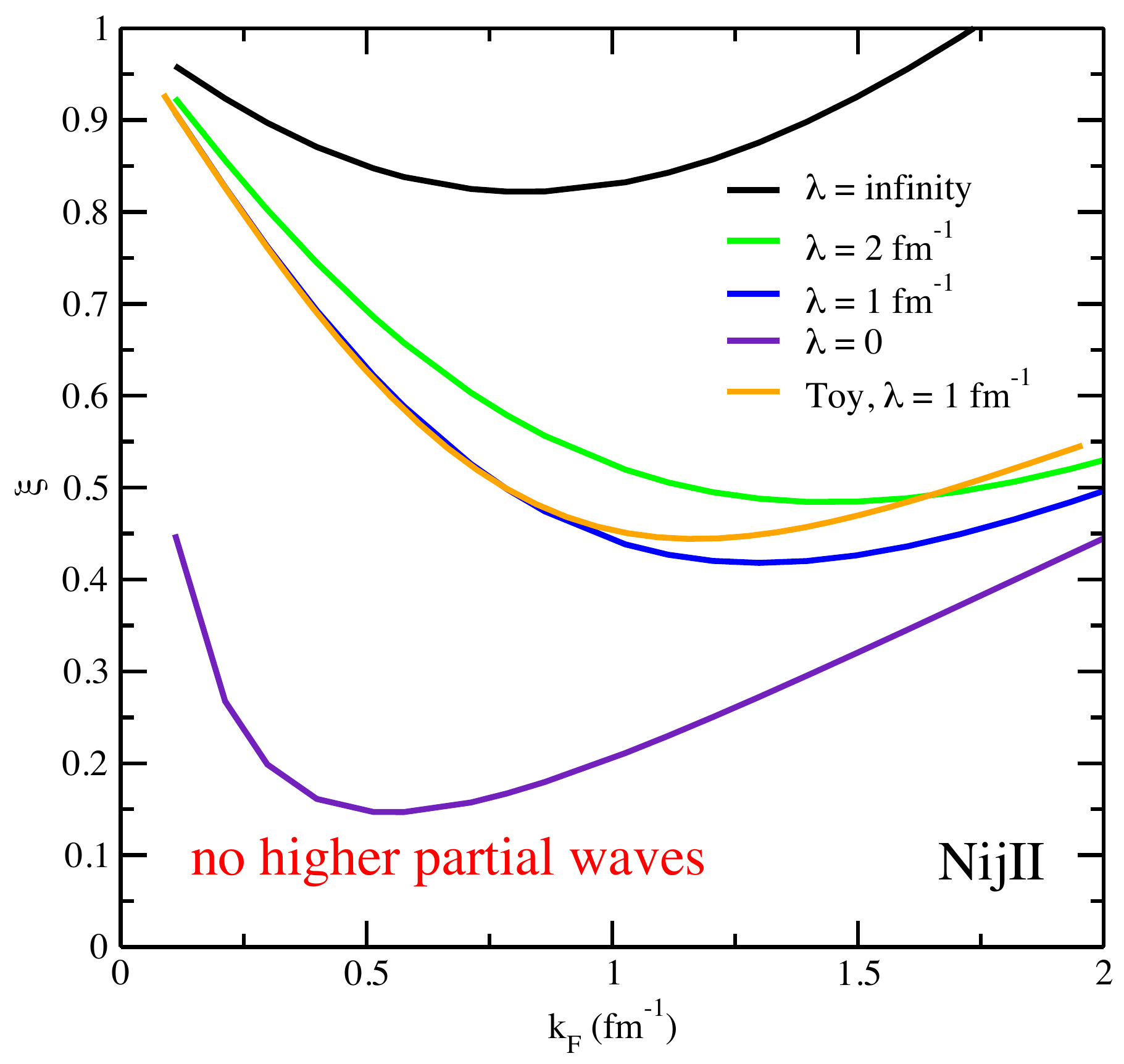} \\ 
\includegraphics[width=12pc]{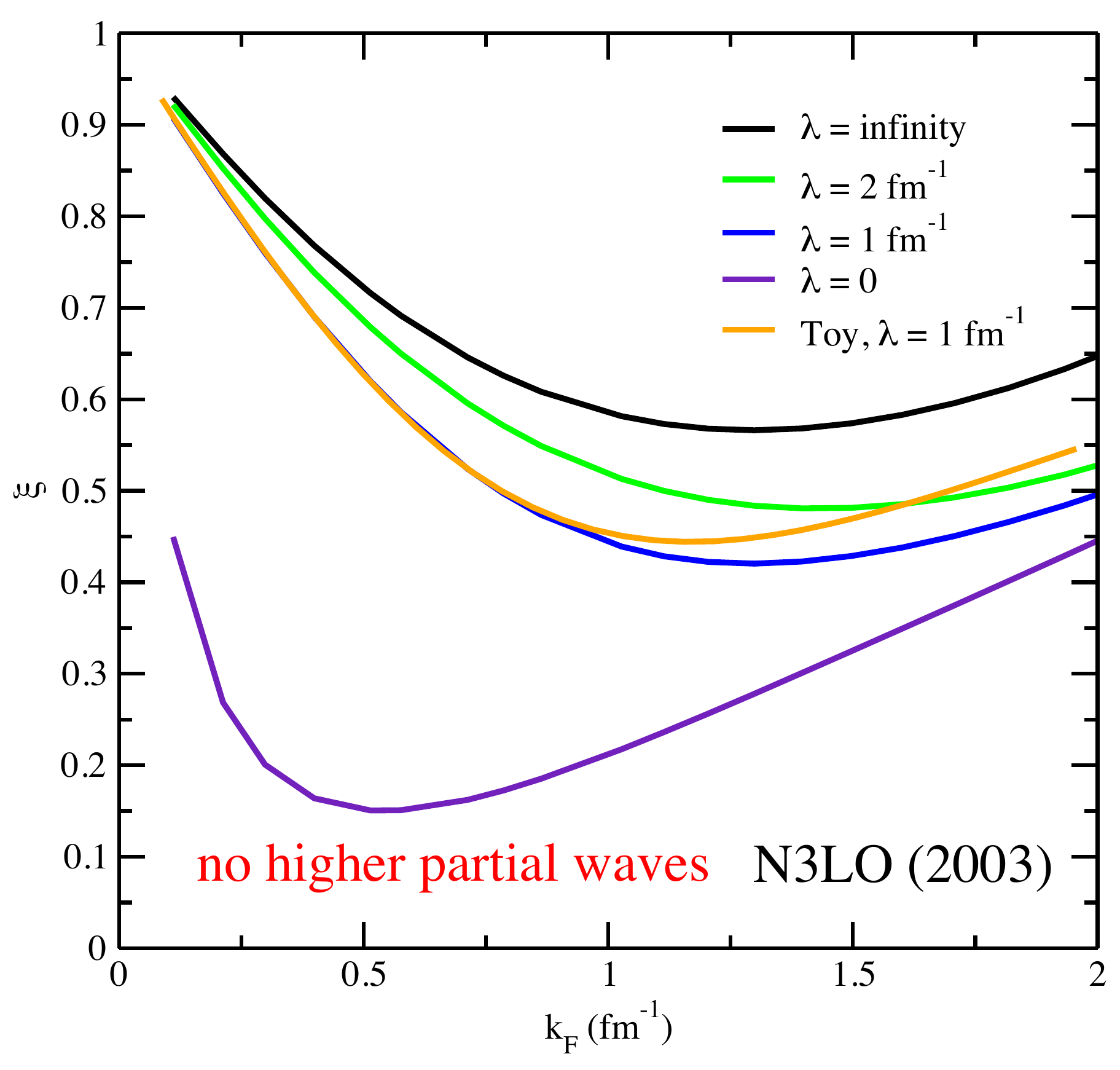}\hspace{1cm}
\includegraphics[width=12pc]{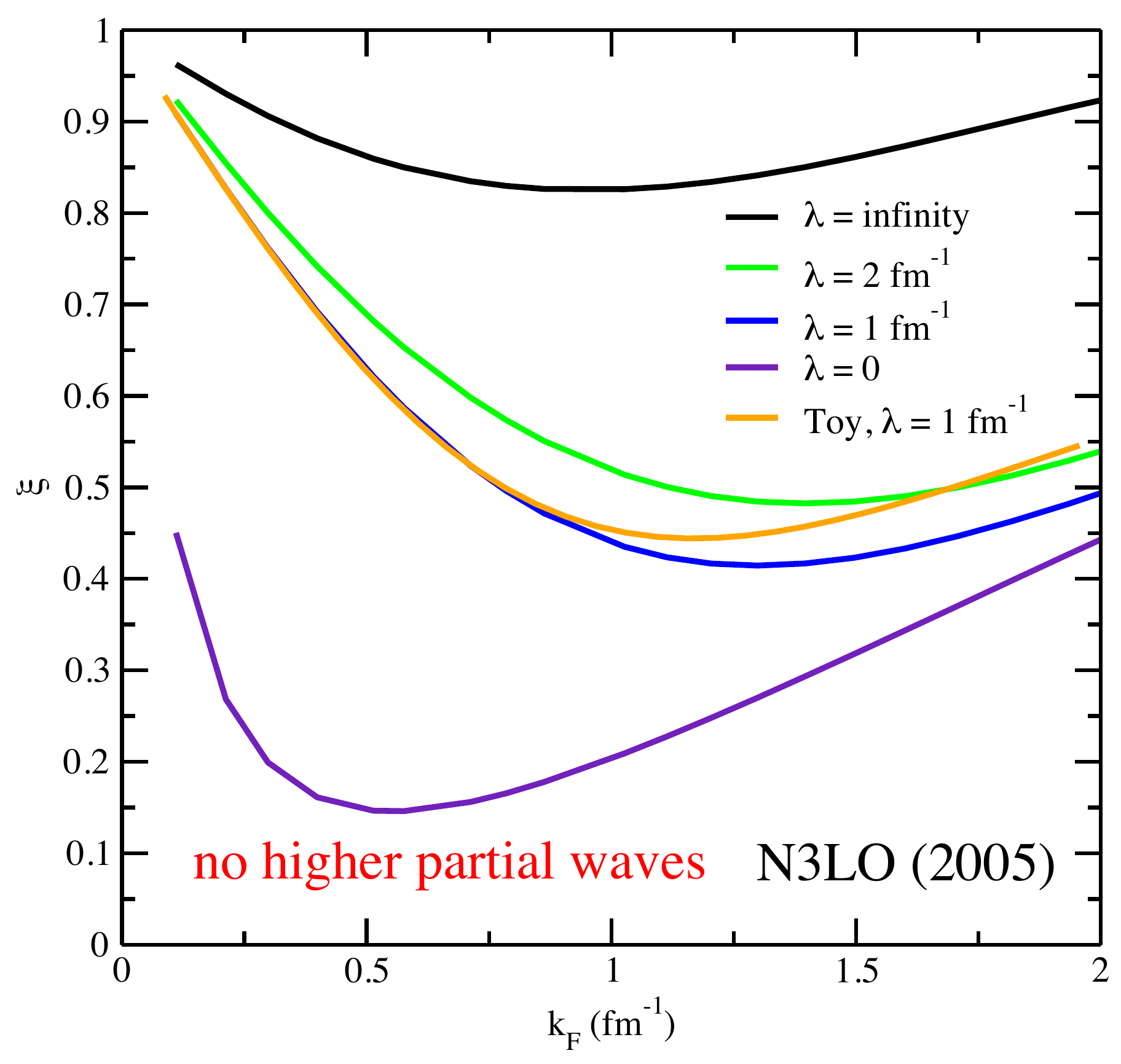} 
\end{center}
\caption{Bertsch parameter from high precision potentials for some values of the similarity cutoff $\lambda$ (Wilson generator) 
with $N=200$ grid points, taking only $S$-waves into account.}
\label{fig3}
\end{figure}

At $\lambda = 1~\rm{fm}^{-1}$ the results obtained with the toy model
are very close to the results that come out if we consider only
$S$-waves from high precision nucleon-nucleon potentials. This can be
observed in Fig. \ref{fig3} where we display the Bertsch parameter
computed with Argonne v18 \cite{av18}, Nijmegen II \cite{nij2}, N3LO
(2003) \cite{n3lo-em} and N3LO (2005) \cite{n3lo-egm} nucleon-nucleon
interactions.  The minimum value of the Bertsch parameter lies in
between $\xi=0.42$ and $\xi=0.45$ for both the toy model and high
precision potentials at $k_F$ between $1.1~{\rm fm}^{-1}$ and
$1.3~{\rm fm}^{-1}$. Sophisticated quantum Monte Carlo calculations
give $\xi=0.44$ \cite{ccps2003}, so it is quite impressive that a
simple separable potential can provide results that are so close to
more accurate approaches.

\begin{figure}[b]
\begin{center}
\includegraphics[width=12pc]{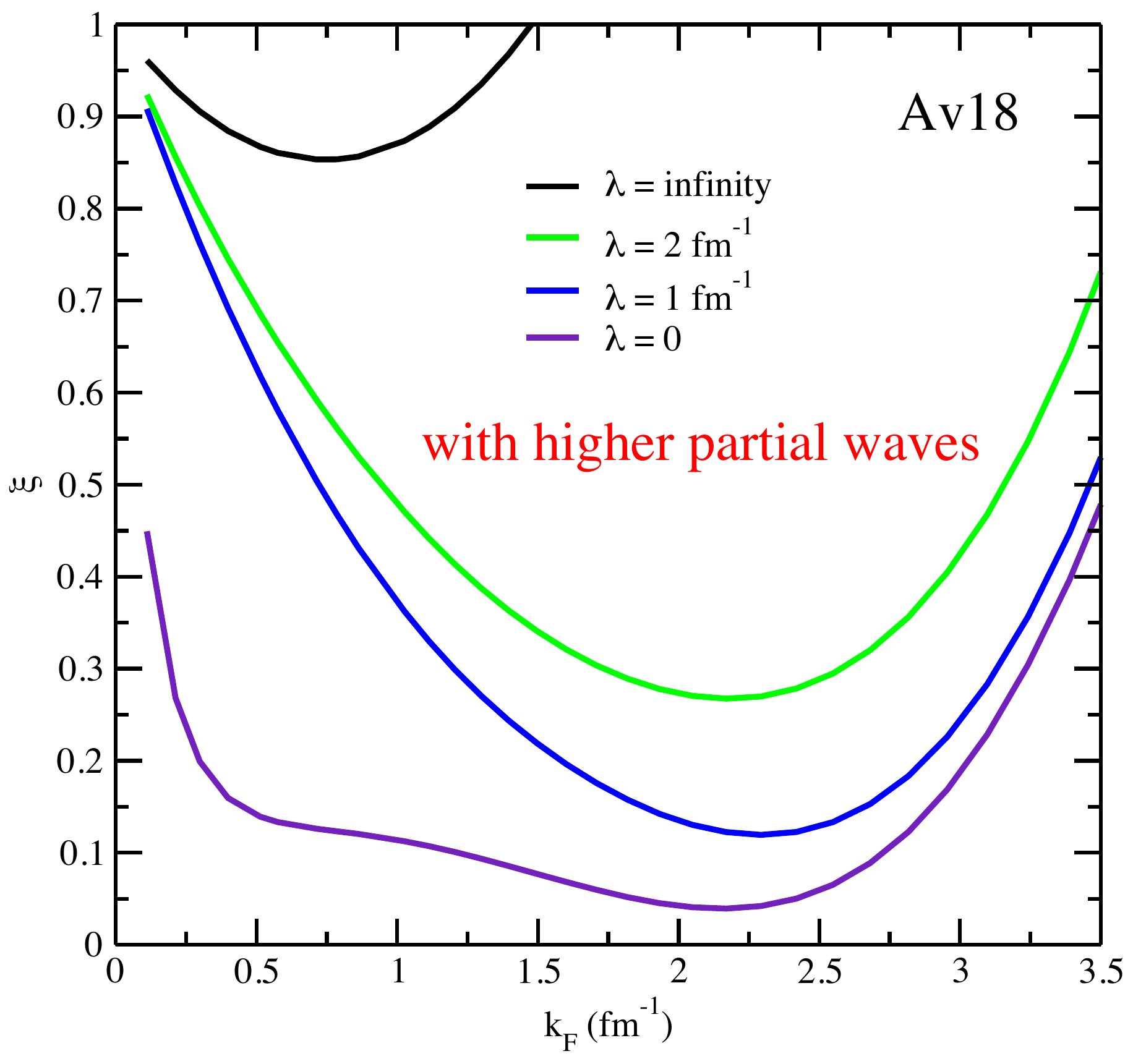}\hspace{1cm}
\includegraphics[width=12pc]{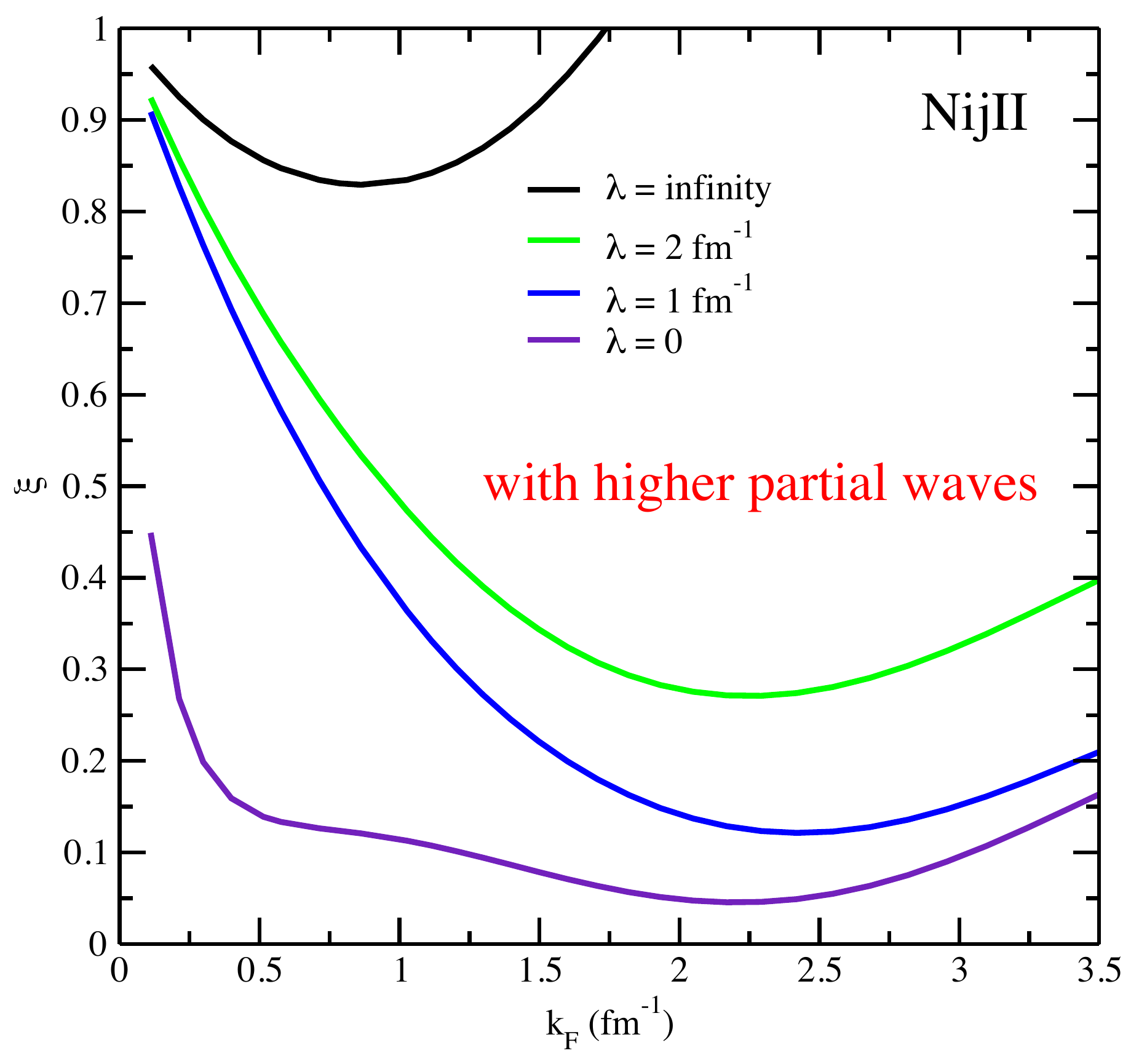} \\ 
\includegraphics[width=12pc]{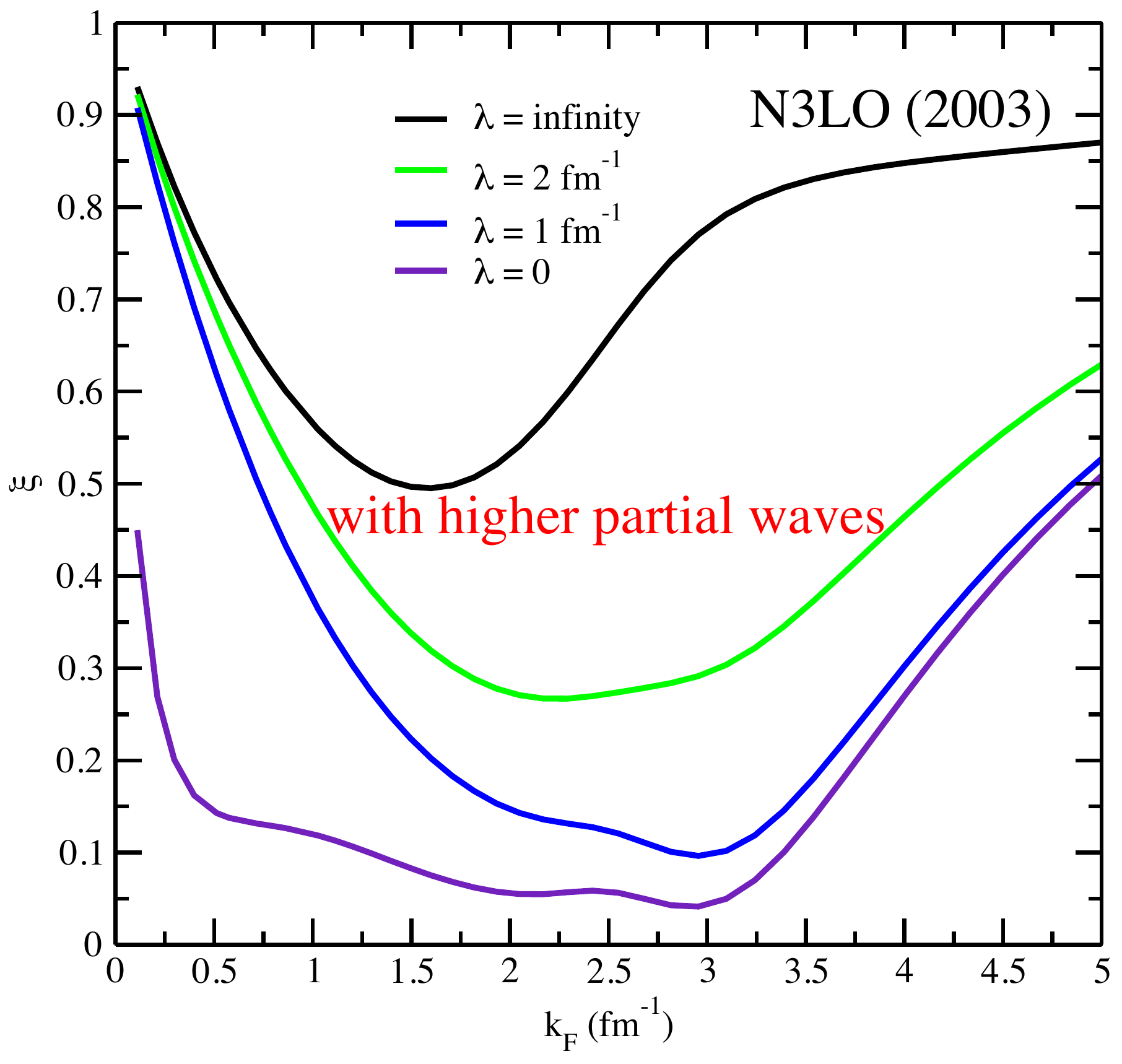}\hspace{1cm}
\includegraphics[width=12pc]{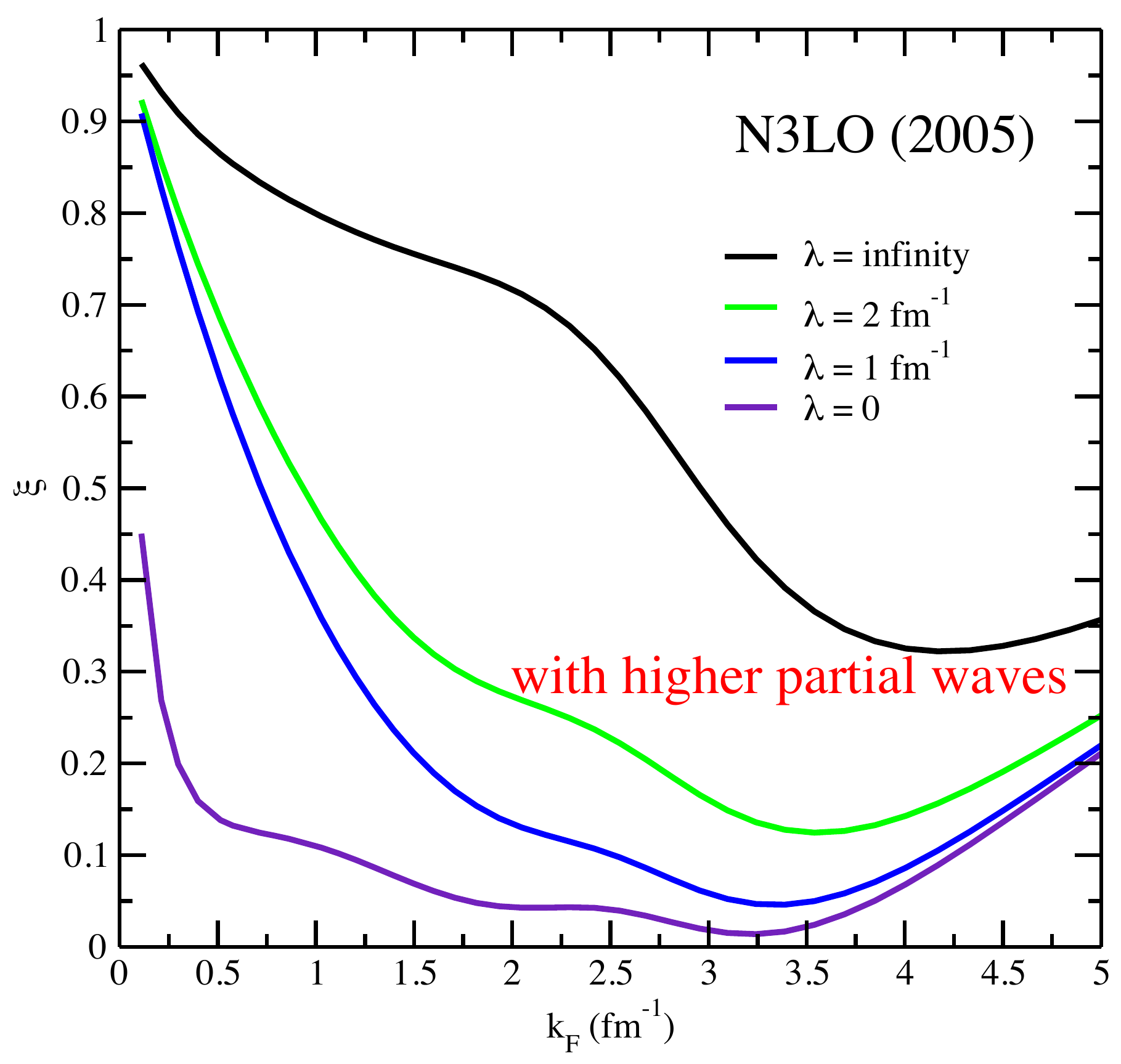} 
\end{center}
\caption{Bertsch parameter from high precision potentials for some values of the similarity cutoff $\lambda$ (Wilson generator) 
with $N=200$ grid points, but summing up to $G$-waves.}
\label{fig4}
\end{figure}

While the contribution from the $S$-waves is rather independent of the
nucleon-nucleon potential, when higher partial waves are included the
results depend on whether one uses phenomenological potentials (Av18
or NijII) or chiral potentials (2003 N3LO or 2005 N3LO). This can be
seen in Fig. \ref{fig4} where the Bertsch parameter is computed
summing up to $G$-waves. Also, the minimum of $\xi$ gets much smaller
and is displaced towards larger $k_F$.

\begin{figure}[h]
\begin{center}
\includegraphics[width=17pc]{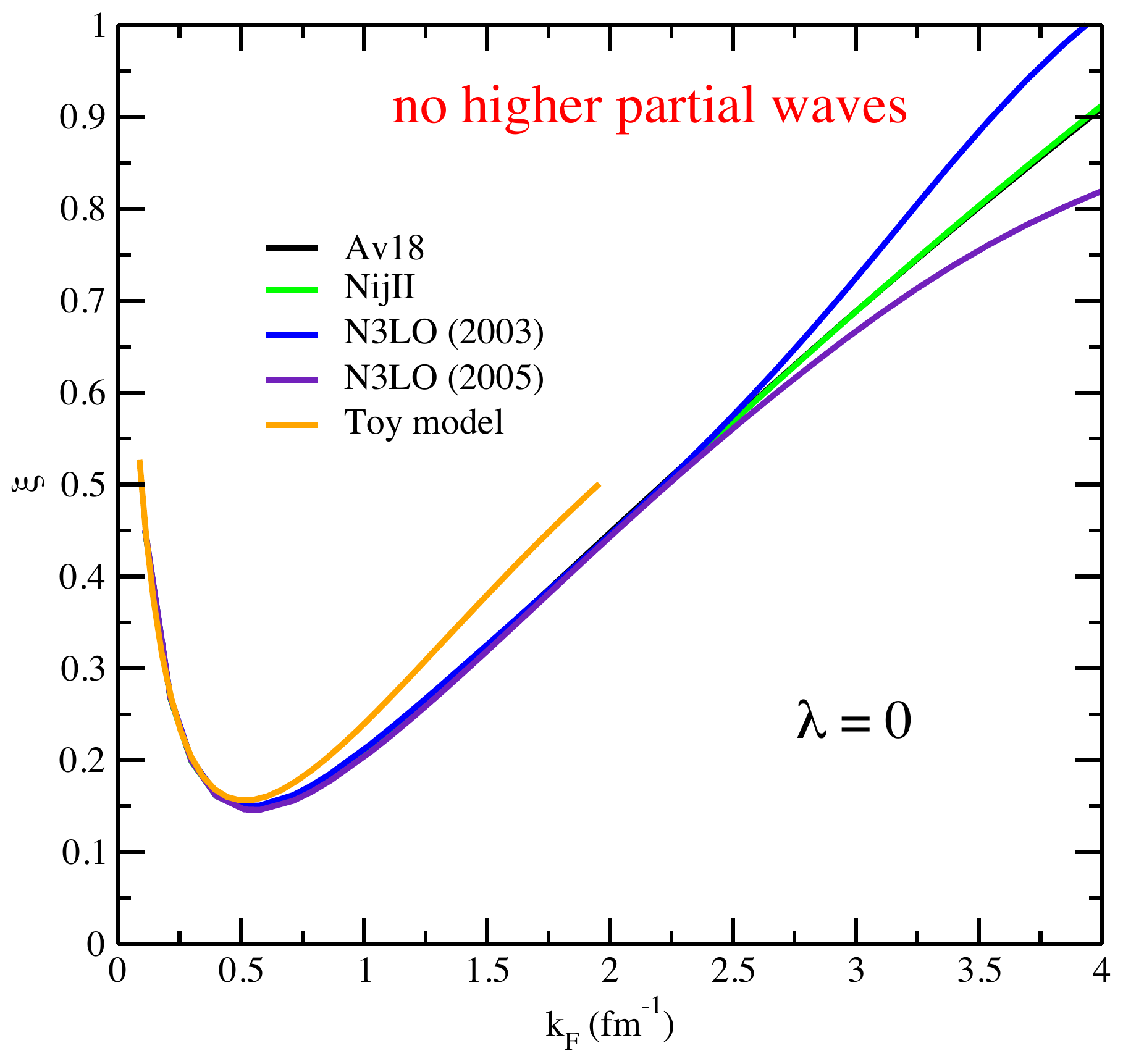} \hspace{0.5cm}
\includegraphics[width=17pc]{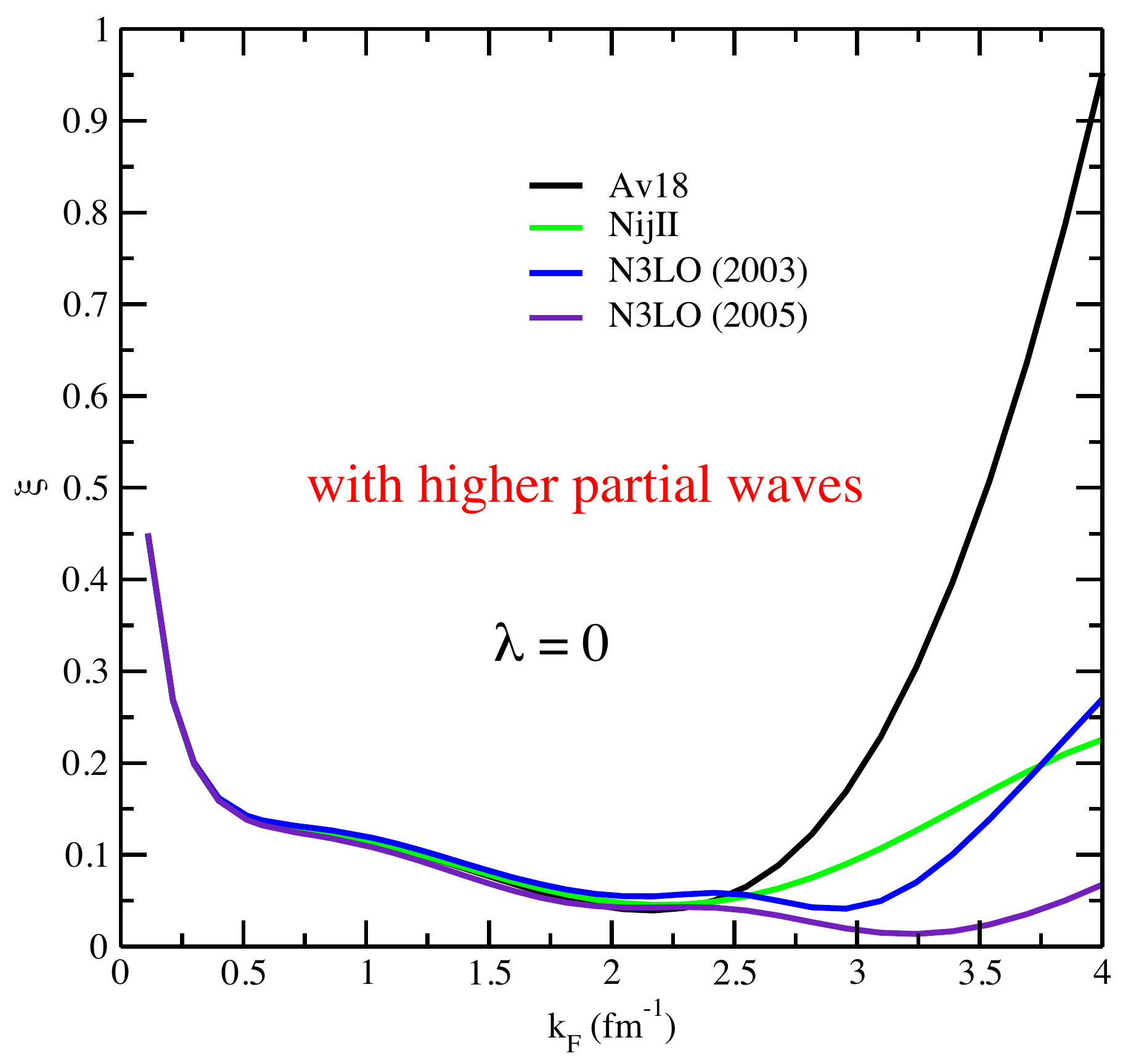} 
\end{center}
\caption{Bertsch parameter in the limit $\lambda=0$ for some high precision potentials with $N=200$ grid points
with only $S$-waves (left) and summing up to $G$-waves (right).}
\label{fig5}
\end{figure}

When the similarity cutoff reaches the limit $\lambda=0$ the
interaction becomes fully diagonal and all off-shell ambiguities are
then eliminated. In Fig. \ref{fig5} we show a comparison of the
Bertsch parameter at $\lambda=0$ for the toy model and the high
precision potentials with only $S$-waves (left panel) and summing up to
$G$-waves (right panel). In the region $0<k_F<0.5~{\rm fm}^{-1}$, the
toy model matches the calculation with high precision potentials with
only $S$-waves.  For larger $k_F$ the result is also very reasonable
if one considers the extreme simplicity of the separable potential.

The $\lambda$-dependence of the Bertsch parameter is rather strong and
shows that it is not determined just by two-body scattering
information unless the extreme on-shell limit $\lambda \to 0$ is
taken. Of course, the real limit $\alpha_0 \to -\infty$ and $\lambda
\to 0$ remains an interesting challenge.

\clearpage


\clearpage

\section{Conclusions}

So far we have studied the ground state of neutron matter in the
unitary limit with a two-nucleon interaction given by a simple
separable potential and compared the results to what is obtained with
high precision nucleon-nucleon interactions.  We also extended the
study to the infrared region of the similarity cutoff by evolving the
interactions with the SRG flow equation for the Wilson generator. The
limit $\lambda=0$ was obtained by using our energy shift
prescription. Our results for the toy model provide a good estimate
for the Bertsch parameter at $\lambda=1~{\rm fm}^{-1}$. In the limit
$\lambda=0$, since the two-body force gets small, $\xi$ lowers
dramatically at intermediate $k_F$.
\section*{Acknowledgements}

E.R.A. would like to thank the Spanish DGI (Grant FIS2011-24149) and
Junta de Andalucia (Grant FQM225) for support. S.S. is partially
supported by FAPESP and V.S.T. thanks FAEPEX, FAPESP and CNPq for
financial support.

\end{document}